\documentclass{article}

\usepackage{amssymb}  
\usepackage{amsmath}
\usepackage{arxiv}
\usepackage[utf8]{inputenc} 
\usepackage[T1]{fontenc}    
\usepackage{hyperref}       
\usepackage[normalsize]{caption} 
\usepackage{floatrow} 
\usepackage{float}
\usepackage[position=bottom, font=Large]{subfig} 
\usepackage{bigdelim}
\usepackage{booktabs} 
\usepackage{hhline} 
\usepackage{longtable}
\usepackage{supertabular}
\usepackage{multirow}
\usepackage{tabu} 
\usepackage{tabularx} 
\usepackage{graphicx}
\usepackage{comment}
\usepackage{mathtools} 

\title{TranSTYLer: Multimodal Behavioral Style Transfer for Facial and Body Gestures Generation}

\date{} 					

\author{
{
\hspace{1mm}Mireille Fares}\\
    ISIR, STMS \\
	Sorbonne University\\
	Paris, France\\
	\And
 {
 \hspace{1mm}Catherine Pelachaud} \\
    ISIR, CNRS \\
	Sorbonne University\\
	Paris, France\\
 \And
  {
  \hspace{1mm}Nicolas Obin}\\
    STMS \\
	Sorbonne University\\
	Paris, France\\}

\renewcommand{\headeright}{ }
\renewcommand{\undertitle}{ }
\renewcommand{\shorttitle}{TranSTYLer}

\hypersetup{
pdftitle={A template for the arxiv style},
pdfsubject={q-bio.NC, q-bio.QM},
pdfauthor={David S.~Hippocampus, Elias D.~Striatum},
pdfkeywords={First keyword, Second keyword, More},
}

\begin{document}

\maketitle
\begin{figure}[h!]
    \centering  
  \includegraphics[width=0.5\textwidth]{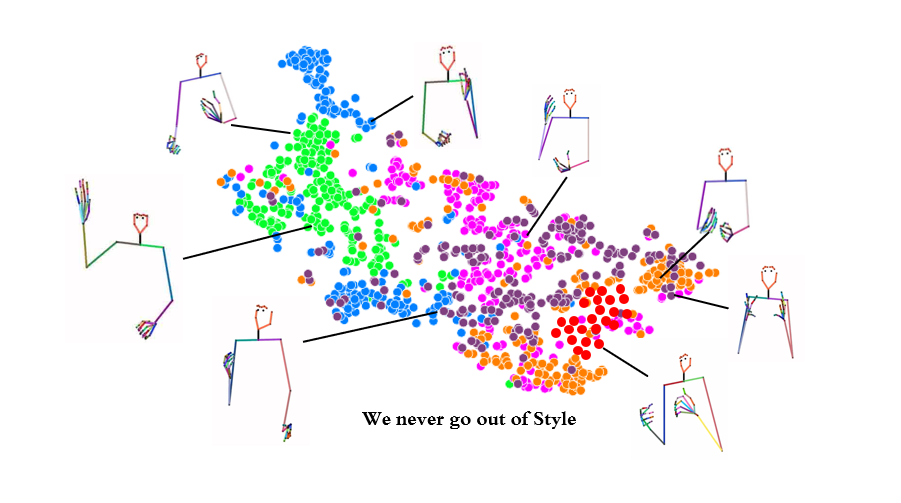}
  \label{fig:teaser}
\end{figure}
\begin{abstract}
This paper addresses the challenge of transferring the behavior expressivity style of a virtual agent to another one while preserving behaviors shape as they carry communicative meaning. Behavior expressivity style is viewed here as the qualitative properties of behaviors. We propose \textit{TranSTYLer}, a multimodal transformer-based model that synthesizes the multimodal behaviors of a source speaker with the style of a target speaker. \textcolor{black}{We assume that behavior expressivity style is encoded across various modalities of communication, including text, speech, body gestures, and facial expressions.} The model employs a style-content disentanglement schema to ensure that the transferred style does not interfere with the meaning conveyed by the source's behaviors. 
Our approach 
eliminates the need for style labels and allows the generalization to styles that have not been seen during the training phase. We train our model on the \textit{PATS corpus}, which we extended to include dialog acts and 2D facial landmarks. Objective and subjective evaluations show that our model outperforms state-of-the-art models in style transfer for both seen and unseen styles during training. 
\textcolor{black}{To tackle the issues of style and content leakage that may arise
, we propose a methodology to assess the degree to which behavior and gestures associated with the target style are successfully transferred, while ensuring the preservation of the ones
related to the source content.}
\end{abstract}

\section{Introduction}
Human communication is a complex phenomenon that involves various modes of expression beyond speech production. It is inherently multimodal, as it relies on the interplay of verbal and non-verbal signals to convey semantic and pragmatic content and facilitate the communication process \cite{knapp2013nonverbal, argyle2013bodily, feyereisen1991gestures, armstrong1995gesture}. \textit{Human behavior expressivity style} refers to the unique and characteristic pattern of behavior exhibited by an individual in various social and communicative contexts \cite{knapp2013nonverbal}. \textcolor{black}{It 
is not a fixed attribute of the speaker but rather is constantly adjusted, accomplished, and co-created with the audience \cite{mendoza1999style}. 
It involves the way an individual communicates verbally and non-verbally, including verbal skills, body language, gestures, and self-expression.} 
\textcolor{black}{Variability in speakers' gesturing is influenced by factors such as their personality traits \cite{mccrae1997personality}, verbal skills \cite{hostetter2007raise}, age \cite{alibali2009gesture, feyereisen1999mental}, and culture \cite{kita2009cross}. The topic and context of the conversation, speaker's role, and relationship with the interlocutor also play a role \cite{hostetter2012effects}. For example, extroverts tend to use larger spatial gestures\cite{hostetter2012effects}.}
\textcolor{black}{Behavior expressivity style can vary between formal and spontaneous speech. In formal settings, a controlled and structured speaking style is used, with formal language and restrained gestures to convey professionalism and respect. In contrast, in spontaneous speech, individuals adopt a more relaxed communication style with informal gestures.} \textcolor{black}{In this paper, we propose a novel machine learning approach to synthesize facial and upper-body gestures driven by \textit{prosodic features} and \textit{text} in the style of different speakers including those unseen during training. We view behavior expressivity style as a pervasive factor during speech, influencing the expressiveness of communicative behaviors, while speech content is conveyed through a combination of multimodal behaviors and text. We propose \textit{TranSTYLer} a transformer-based model that can synthesize facial and body gestures of a source speaker in the style of any target speaker, while ensuring that the transferred style does not interfere with the meaning conveyed by the source gestures. Our approach incorporates a disentanglement scheme that separates content and style, enabling us to directly infer the style representation even for speakers who were not part of the training process, without requiring additional training or fine-tuning. Our system comprises two main components. Firstly, we have a speaker style encoder network, which learns to generate a fixed-dimensional embedding that represents the style of a target speaker. This embedding is derived from the target speaker's multimodal data (facial and body gestures, audio, and text). Secondly, we employ a synthesis network that synthesizes gestures based on the content provided by the input modalities (text and audio) of a source speaker. This synthesis process is conditioned on the target speaker style embedding, ensuring that the generated gestures exhibit the target style characteristics. We also introduce a new methodology to measure the transferred style and the preservation of gestures that convey meaning. }
\textcolor{black}{We evaluate 
the performance of \textit{TranSTYLer} in terms of \textit{style transfer} and \textit{content preservation}. Objective and subjective evaluations confirm the quality of our approach, outperforming two state-of-the-art models.
This paper is organized as follows. We start by providing a review of the existing \textit{behavior expressivity style} modeling approaches, discussing their limitations. We then explain our contributions and the architecture we propose. Finally, we present objective and subjective evaluations and discuss the results.}
\begin{figure*}[t]
\includegraphics[width=0.7\linewidth]{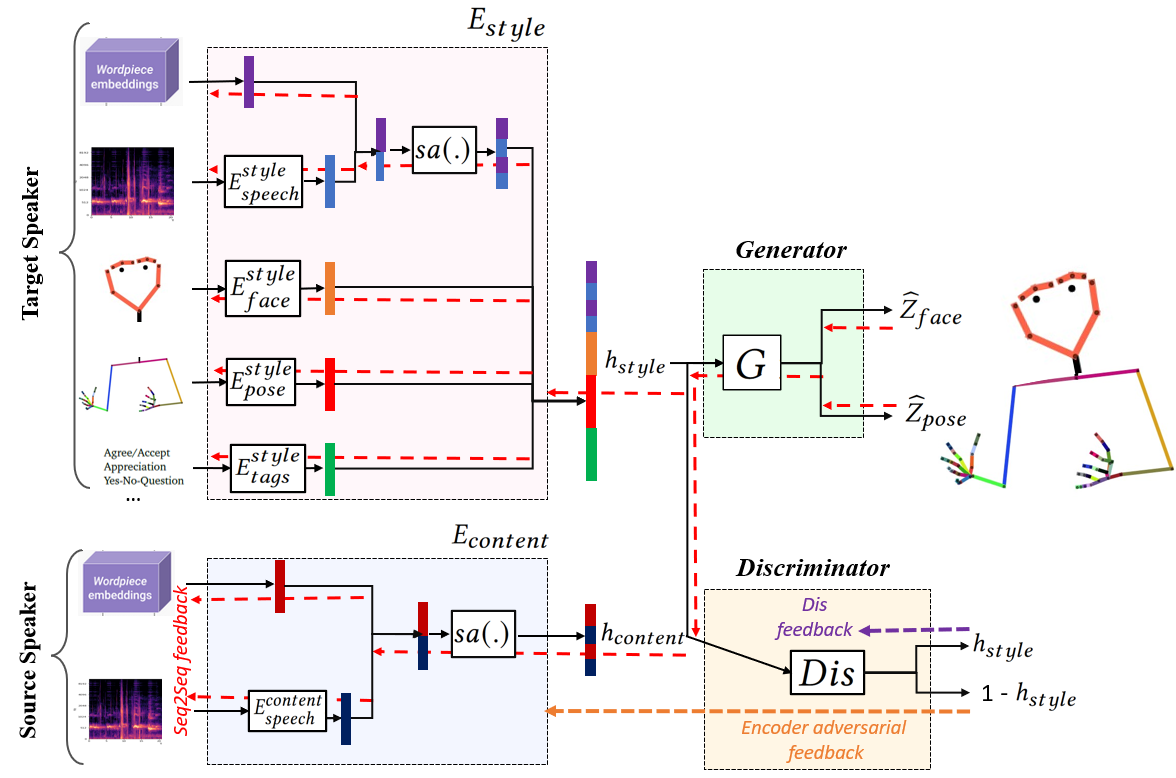}
\caption{Overview of \textit{TranSTYler}, an approach 
driven by the content of a source speaker's speech - text semantics ($X_{text}^{source}$) and Mel spectrogram ($X_{speech}^{source}$) - and conditioned on a target speaker's style vector $h_{style}$ generated from the target's multimodal input data - Mel spectrogram ($X_{speech}^{target}$), 2D facial landmarks ($X_{face}^{target}$), 2D poses ($X_{pose}^{target}$), and dialog tags ($X_{tags}^{target}$). The network is composed of: (1) a \textit{style encoder} that encodes the target's multimodal data and generates the style vector $h_{style}$, (2) a \textit{content encoder} that encodes the source's speech content and generates the content vector $h_{content}$, and (3) a \textit{discriminator} $Dis$ used during training time to disentangle $h_{content}$ and $h_{style}$. 
A generator $G$ is used to generate body $\widehat{Z}_{pose}$ and facial $\widehat{Z}_{face}$ gestures.}
\label{ZS-MSTM2.0_Architecture}
\end{figure*}
\section{Related Works}
\textcolor{black}{
Gesture style modeling and control have gained significance in proposing expressive behaviors for virtual agents that can be adjusted and tailored to specific audiences or interlocutors.}
A first model to generate communicative behaviors with different styles was proposed by Neff et al.\cite{neff2008gesture}.
The authors developed a system that generates full body gesture animation based on text, mimicking the style of a specific performer. They focused mainly on hand movements. In recent years, Alexanderson et al. \cite{alexanderson2020style} proposed a generative model that synthesizes speech-driven gesticulation while exerting control over the output style, such as gesture level and speed. Karras et al. \cite{karras2017audio} created a model that generates 3D facial animation from audio to  capture the style of a single actor. Similarly, Cudeiro et al. \cite{cudeiro2019capture} developed a model, called \textit{VOCA}, that synthesizes 3D facial animation driven by speech signal, allowing for a wide range of speaking styles to be realistically animated, even in languages other than English. On the other hand, Ginosar et al. \cite{ginosar2019learning} proposed an approach for generating gestures using models trained on individual speakers.
Yoon et al \cite{yoon2020speech} developed a method to generate gestures that matched a speaker's style by using their speaker identity. Their approach involved processing the input audio and text with separate audio and text encoders and using the speaker identity to select a style embedding from a learned space. These features were then fed to a gesture generator to produce a sequence of poses that matched the content and rhythm of the speech. On the other hand, Ghorbani et al. \cite{ghorbani2022exemplar, ghorbani2022zeroeggs} proposed a framework that improved on high-level style portrayal by using exemplar motion sequences to demonstrate the intended stylistic expression of gesture motion. Their method could extract style parameters in a zero-shot manner, only requiring a single example motion and was able to generalize to example motions (and therefore styles) not seen during training.
The works just mentioned have focused on generating \textcolor{black}{behaviors from one modality, either facial expressions, head movements or gestures. }
However, they have not considered multimodal data when modeling style or synthesizing gestures. Additionally, their generative models were only trained on \textit{single-speaker} data. \textcolor{black}{Recently, Ahuja et al. \cite{ahuja2020style} have attempted to model and transfer style from a \textit{multi-speaker database}. They proposed \textit{Mix-StAGE}, a speech-driven approach that trains a model using data from multiple speakers while learning a unique style embedding for each speaker. Their neural architecture uses a \textit{content} and \textit{style} encoder to extract content and style information from pose. A style embedding matrix is used to represent the style associated with each specific speaker in the training set. Their approach presents several limitations. First, behavior expressivity style is only encoded by means of upper-body motion, ignoring the other possible modes of style expression, such as text, speech, and facial expresions. Second, speakers are associated with a unique speaker identity "ID", considered as style labels, which hinder their ability to generalize to new speakers. Later on, Ahuja et al. \cite{Ahuja_CVPR_lowRes} presented a few-shot style transfer strategy based on neural domain adaptation to transfer style with only a few examples, considering the shift in cross-modal grounding between the source speaker and the target style. However, this approach still requires to have at least 2 minutes of the style to be transferred.}
\textcolor{black}{Fares et. al \cite{fares2022zero, fares2023zero} proposed an approach to synthesize upper body gestures of a source speaker in the style of any target speaker. The authors do not consider faces in their model. Their approach can be applied to speakers whose style behaviors have been learned or not during the training phase.}
\textcolor{black}{Overall,} the recent models proposed to capture behavior expressivity style have several limitations: they do not exploit multimodal data \cite{ye2022audio, neff2008gesture, alexanderson2020style, karras2017audio, cudeiro2019capture, ginosar2019learning, Ginosar_2019_CVPR, fares2022zero}; their generative models are trained on single speaker data; style is associated with speaker "IDs", which limits their ability to generalize to new speakers without additional training \cite{cudeiro2019capture, karras2017audio, ginosar2019learning}; they require additional training to model unseen target speaker style \cite{Ahuja_CVPR_lowRes}. In addition, they lack a methodology to evaluate properly the behavior expressivity style transfer. In particular, during the style transfer from a target style to a source content, the resulting behavior should ideally preserve the gesture related to the source content (e.g., idiomatic gestures) while modifying its expression accordingly to the target style. Practically, it is a common pitfall in style transfer to observe leakage between the source content and the target style, i.e. partially preserving the source style or transferring the target content. 
\section{Our contributions}
To address the limitations, we introduce \textit{TranSTYLer} a transformer-based model for the generating facial and body gestures of a source speaker in the style of a target speaker, while preserving the intended meaning of the source gestures. Our contributions are:
\begin{enumerate}
    \item To the best of our knowledge, \textit{TranSTYLer} is the first behavior expressivity style transfer approach that jointly synthesizes \textit{2D upper-body gestures} and \textit{2D facial landmarks} of \textit{source speakers}, in the style of {any} \textit{target speakers}, and without requiring additional training, making our approach zero-shot. 
    \item  We propose a novel methodology for assessing \textit{behavior expressivity style transfer} for generating communicative behaviors for virtual agents. Our methodology measures \textit{content preservation} and \textit{style transfer} and gives insights about potential leakages between style and content information.
    \item We have extended the \textit{PATS corpus}, by including \textit{2D Facial Landmarks} and \textit{Dialog Tags}.
\end{enumerate}
\section{\textit{TranSTYler}}
We present \textit{\textbf{TranSTYler}}, a novel approach for modelling \textit{behavior expressivity style} in virtual agents.\textit{\textbf{TranSTYler}} is a multimodal style transfer approach for generating 2D facial and pose synthesis corresponding to the \textit{content} of a \textit{source speaker} and in the \textit{style} of a \textit{target speaker}. During inference, an embedding style vector can be directly inferred from any target speaker's multimodal data - text, speech, poses, 2D facial landmarks - by simple projection into \textit{\textbf{TranSTYler}}'s embedding style space (similar to the one used in \cite{jia2018transfer} and \cite{fares2023zero}). Our approach allows for style transfer from any unseen speakers, without requiring further training or fine-tuning of our trained model. This means that our method is not restricted to the styles of the speakers in the training dataset. \textcolor{black}{\textit{\textbf{TranSTYler} }is trained on PATS corpus \cite{ahuja2020style} which we extended to include additional facial and text features.}
\subsection{Problem Positioning}
\textcolor{black}{The goal is to learn to generate facial and upper-body gestures 
based on the source speaker's content information 
and conditioned on the style information 
of the target speaker. A transformer-based generator 
is used to generate  facial and body gestures from content and style information. 
An adversarial component in the form of a fader network \cite{lample2017fader} is used for disentangling style and content from the multimodal data. At inference time, it is discarded, and the model can generate different versions of facial and body gestures when fed with different style vectors. Gesture styles for the same input speech can be directly controlled by switching the value of style vector 
or by calculating it from a target speaker's multimodal data fed as input to the style encoder. 
}
Our approach is based on the following hypotheses: 
\begin{itemize}    
\item \textcolor{black}{Our primary hypothesis is that behavior involves the modulation of communicative gestures associated with content, through the specific gestures associated with an individual. We propose to disentangle this information and encode it in a differentiated manner.}
\item \textit{Behavior expressivity style} is encoded across \textit{text semantics}, \textit{dialog tags}, \textit{speech}, \textit{face} and \textit{pose} and varies little or not over time. The reason we consider \textit{dialog tags} is to capture further semantic information. Moreover, studies on communicative gestures have shown the link between the meaning carried by dialog acts and the one carried by gestures (\cite{Calbris, cienki2005image}). 
\item \textit{Speech content} is encoded across the verbal and nonverbal modalities. That is, the meaning of what is being said is conveyed by the text and by the nonverbal communicative behaviors.
\end{itemize}
To implement theses assumptions, we propose an architecture for encoding and disentangling the source speaker's \emph{content} and the target speaker's \emph{style} information from multiple modalities. \textcolor{black}{Style and content information are entangled in each utterance produced by a speaker.} On one side, a content encoder $E_{content}$ is used to encode a content matrix from the source's text and speech signal; on the other hand, a style encoder $E_{style}$ is used to encode a style vector from the text, acoustic features, dialog tags, facial and body gestures data of the target. A fader loss (\cite{lample2017fader}) is introduced to effectively disentangle content and style encodings. 
The network processes source and target input data at the segment level, where each segment \emph{\textbf{S}} consists of $64$ frames. For each segment \emph{\textbf{S}}, the network takes as input: 
\begin{enumerate}
    \item The source speaker's audio and text semantics represented by the Mel spectrogram ($X_{speech}^{source}$) and Bert embeddings ($X_{text}^{source}$).
    \item The target speaker's facial gestures, body gestures, audio, text and dialog tags represented by 2D facial landmarks ($X_{face}^{target}$), 2D poses ($X_{body}^{target}$), Mel spectrogram ($X_{speech}^{target}$), Bert embeddings ($X_{text}^{target}$), and dialog tags ($X_{tags}^{target}$).
\end{enumerate}
For each \textit{\textbf{S}}, the output of the network is the generation of behaviors that correspond to the content of the source speaker with the style of the target speaker, namely:
\begin{enumerate}
    \item Facial gestures ($\widehat{Z}_{face}$) represented by 2D facial landmarks.
    \item The corresponding upper-body gestures represented by 2D poses ($\widehat{Z}_{body}$).
\end{enumerate}

\subsection{Neural Formulation}
The network has an embedding size $d_{model}$ equals to $64$.
\\[0.2cm]
\textbf{Content Encoder. }
$E_{content}$ \textcolor{black}{encodes the source speaker's speech content information from the variables } $X_{text}^{source}$ and $X_{speech}^{source}$ corresponding to each \textbf{S}. $X_{text}^{source}$ is represented by BERT embeddings of dimension $768$. $X_{speech}^{source}$ is encoded using $E_{speech}^{content}$, a pre-trained \emph{Mel Spectrogram Transformer (AST)} \emph{base384} model (\cite{gong2021ast}), and then concatenated with $X_{text}^{source}$. 
A self-attention mechanism is then applied on the resulting vector. The multi-head attention layer has $N_{h}$ equals to 4 attention heads, and an embedding size $d_{att}$ equals to $d_{att}=d_{model}+768$. The output of the attention layer is the vector $h_{content}$, a content representation of the source speaker's speech, which can be written as follows:
\begin{equation} \label{eqn1}
	h_{content} =  sa\left( \left[ E_{speech}^{content} (X_{speech}^{source}),  X_{text}^{source} \right] \right)
\end{equation}
where: sa(.) denotes self-attention.
\\[0.2cm]
\textbf{Style Encoder.} We consider that \emph{behavior expressivity style} is encoded in a speaker's multimodal behavior. As illustrated in Figure \ref{ZS-MSTM2.0_Architecture}, \textcolor{black}{$E_{style}$ encodes the behavior expressivity style information from the target speaker's variables }
$X_{speech}^{target}$, $X_{text}^{target}$, $X_{pose}^{target}$, $X_{face}^{target}$, and $X_{tags}^{target}$; which are encoded by $E_{speech}^{style}$, $E_{text}^{style}$, $E_{pose}^{style}$, $E_{face}^{style}$, and  $E_{tags}^{style}$ respectively. $E_{pose}^{style}$ and $E_{face}^{style}$ are composed of $N_{lay}$ equals to 3 layers of LSTMs with a hidden-size equal to $d_{model}$. $E_{tags}^{style}$ is a \textit{One Hot Encoder} that considers the 38 dialog tags as categorical features. The input features are encoded using a one-hot encoding scheme. The output is a sparse array containing binary values representing the presence or the absence of each tag in the segment \textbf{\textbf{S}}. $X_{speech}^{target}$ is encoded by $E_{speech}^{style}$, which is the pre-trained \emph{AST}. The output vector is concatenated with $X_{text}$ and a self attention mechanism is applied on the resulting vector. This attention layer has $N_{h}$ equals to 4 attention heads and an embedding size equals to $d_{att}$. Finally, the output vector is concatenated with the other encoded modalities. The resulting vector $h_{style}$  is the output speaker style embedding that serves to condition the network with the speaker style. The final style embedding $h_{style}$ can therefore be written as follows:
\begin{equation} \label{eq2}
\begin{split}
 		h_{style} = \bigl[sa \left( \left[X_{text}^{target}, E_{speech}^{style}  (X_{speech}^{target}) \right] \right),  \\
  E_{pose}^{style} (X_{pose}^{target}),   E_{face}^{style} (X_{face}^{target}), E_{tags}^{style} (X_{tags}^{target})\bigr]   
\end{split}
\end{equation}
where: sa(.) denotes self-attention.

\textbf{Generator. } 
For decoding $\widehat{\textit{Z}}_{\textbf{\textit{pose}}}$, and $\widehat{\textit{Z}}_{\textbf{\textit{face}}}$, the sequence $h_{content}$ and the vector $h_{style}$ are concatenated (by repeating the $h_{style}$ vector for each segment of the sequence), and passed through a $Dense$ layer of size $d_{model}$. 
We then give the resulting vector as input to two \textit{Transformer Decoders}. Each \textit{Transformer Decoder} is composed of $N_{dec}$\emph{ = 1} decoding layer, with $N_{h}$ = 2 attention heads, and an  embedding size equal to $d_{model}$. 
The resulting output vectors are sequences of 2D facial landmarks and 2D-poses which corresponds to: 
\begin{equation}\label{loss_zsmstm1.0pose}
\begin{split}
\widehat{Z}_{pose} = G_{pose}(h_{content}, h_{style})\\
\widehat{Z}_{face} = G_{face}(h_{content}, h_{style})
\end{split}
\end{equation}
where $G_{face}$ and $G_{pose}$ are the transformer decoders corresponding to \textit{face} and \textit{pose} modalities. 
The generator loss can therefore be written as:
\begin{equation} \label{eqn3}
\begin{split}
    	\mathcal{L}^{gen}_{rec}(E_{content}, E_{style}, G) = \mathbb{E}_{\widehat{Z}_{pose}} ||{Z}_{pose} - G_{pose}(h_{content}, h_{style}) ||_2 
     \\
     + \mathbb{E}_{\widehat{Z}_{face}} ||{Z}_{face} - G_{face}(h_{content}, h_{style}) ||_2
\end{split}
\end{equation}

\textbf{Adversarial Component. }
Our approach of disentangling \textit{style} from \textit{content} relies on the fader network disentangling approach (\cite{lample2017fader}), where a fader loss is introduced to effectively separate $h_{style}$ and $h_{content}$. The latent space of $h_{content}$ is constrained to be independent of $h_{style}$ embeddings. Concretely, it means that the distribution over $h_{content}$ of the latent representations should not contain the style information. 
We formulate this discriminator $Dis$ as a regression on the conditional variable $h_{style}$. $Dis$ learns to predict $h_{style}$ from  $h_{content}$, as:
\begin{equation}
\widehat{h}_{style} = Dis({h}_{content})
\end{equation}
While optimizing the discriminator, the discriminator loss $\mathcal{L}^{dis}$ must be as low as possible, such as:
\begin{equation}
\mathcal{L}^{dis}(Dis) = \mathbb{E}_{\widehat{h}_{style}} ||{h}_{style} - Dis(h_{content}) ||_2
\end{equation}
In turn, while optimizing the generator loss including the fader loss $\mathcal{L}^{gen}_{adv}$, the discriminator must not be able to predict correctly $h_{style}$ from $h_{content}$ conducting to a high discriminator error and thus a low fader loss. The adversarial loss can be written as:
\begin{equation}
\mathcal{L}^{gen}_{adv}(E_{content}, E_{style}) = \mathbb{E}_{{h}_{style}} || 1 - (h_{style} - Dis(h_{content})) ||_2
\end{equation}
The style prediction error is preliminary normalized within 0 and 1 range. The total $G$ loss can therefore be written as follows:
\begin{equation} \label{eqn4}
\begin{split}
	\mathcal{L}^{gen}_{total}(E_{content}, E_{style}, G)  = \mathcal{L}^{gen}_{rec}(E_{content}, E_{style}, G) + \\
\lambda \mathcal{L}^{gen}_{adv}(E_{content}, E_{style}, G)
 \end{split}
\end{equation}
where $\lambda$ is the adversarial weight that starts off at 0 and is linearly incremented by 0.01 after each training step. The discriminator $Dis$ and the generator $G$ are then optimized alternatively as described in \cite{lample2017fader}. All \textbf{\textit{\textit{TranSTYler}}} hyperparameters were chosen empirically and are listed \textcolor{black}{in the implementation details section of the appendix.}
\section{Experimental Evaluations}
\subsection{Material and Model setups}
\textbf{PATS 2.0 Corpus. }The \emph{PATS Corpus} \cite{ahuja2020style, ginosar2019learning} originally includes \textit{2D upper-body joints keypoints}, aligned with the given speech, \textit{Mel spectrogram} and  \textit{Bert embeddings}, of 25 speakers, categorized as follows: 15 talk show hosts, 5 lecturers, 3 YouTubers, and 2 televangelists. Each speaker has his/her own communicative style, and lexical and gesture diversity. It has $251$ hours of data, with $84,000$ intervals and a mean duration equal to  $10.7$ seconds per interval. The standard deviation is $13.5$ seconds per interval. An interval corresponds to an utterance consisting of 64 timesteps. We have extended PATS to include 
\emph{2D facial landmarks},
and \emph{dialog tags}, 
More specifically, we extracted 70 2D facial landmarks for all PATS speakers using OpenPose \cite{cao2017realtime} and aligned with PATS's features. 
Dialog acts correspond to the communicative functions expressed by the spoken text \cite{bunt2010towards}. 
We used the tool "DialogTag" \cite{bhavitvyamalik} to extract $38$ dialog tags from PATS utterances. 
We refer the readers to the supplementary materials for the complete list of dialog tags. 
\\[0.2cm]
\textbf{TranSTYler Training and Testing. }We trained our network using \textbf{\textit{PATS 2.0}}. Although fingers are included in PATS, we have chosen not to model finger data in our work. The quality of the extracted fingers data is very noisy and lacks accuracy. Instead, we focus on modeling and predicting the 2D joints of the upper body and arms, using 11 joints to represent these areas. We also model 15 facial landmarks, which are illustrated in our Appendix. We use less keypoints than those originally extracted to have less input parameters and fasten the training phase. 
We took out some keypoints from the face contour and 2 keypoints on each eyebrow, and we used only 2 keypoints for the eyelids. 
In total, we model 11 body and arm joints, and 15 facial landmarks. 
An utterance is associated to one or more dialog acts. We consider all the 38 different tags that are listed in our appendix. 
Our testing comprises two conditions: \textit{Seen Speaker} and \textit{Unseen Speaker}. The \textit{Seen Speaker} condition evaluates how accurately our model can perform style transfer when presented with target speakers seen during training. In contrast, the \textit{Unseen Speaker} condition evaluates our model's ability to perform zero-shot style transfer when presented with target speakers that were not seen during training. We carefully selected both seen and unseen speakers from PATS to cover a range of stylistic behaviors in terms of lexical diversity and spatial extent \textcolor{black}{which is the amplitude of body movements.} 
The PATS database already defines the train, validation, and test sets for each speaker. We train our model on 16 PATS speakers. To test the \textit{Seen Speaker} condition, we use the test sets of the 16 PATS speakers as our test set. For the \textit{Unseen Speaker} condition, we select 6 other speakers and use their test sets for our experiments. 
Each training batch has $BS$ = 24 pairs of word embeddings, Mel spectrogram, dialog acts, and their corresponding sequence of (x, y) joints of the skeleton of the upper-body pose and 2D facial landmarks. We use Adam optimizer with $\beta_{1} = 0.95, \beta_{2} = 0.999$, and a \textit{Cyclical Learning Rate} (CLR) scheduler to render the learning balanced. The initial learning rate $Lr_{init}$ of the CLR is equal to $1e-7$, the end learning rate $Lr_{e}$ is equal to $0.1$, and the step size $St_{size}$ is equal to $196$. We train the network for $N_{it}$ equals to $78,400$ iterations. All features values are normalized so that the dataset mean and standard deviation are $0$ and $0.5$, respectively. All hyperparameters used for training are summarized in our appendix.
\subsection{Objective Evaluation}
We objectively measure the performance of \textit{\textbf{TranSTYler}} in terms of two key aspects: \textit{style transfer accuracy} and \textit{content preservation}.
Moreover, to measure the degree of similarity of the generated facial and body gestures with the source and target styles, we computed the distance between our model's predictions and each of the source and target styles. Additionally, \textcolor{black}{we assess and compare the unique dynamic movement patterns of the source, target, and prediction by measuring their velocity, acceleration, and jerk. This allows us to quantify and analyze the specific characteristics of their movement dynamics.} 
\\[0.2cm]
\textbf{Metrics. }We have followed the recommendations put forth by Fu et al. \cite{fu2018style} in order to evaluate the characteristics of style transfer in our study. We employed their proposed evaluation metrics, \textit{Transfer Strength Accuracy} and \textit{Content Preservation}, to assess the performance of \textit{\textbf{TranSTYler}}. These metrics measure quantitatively the accuracy of style transfer and the extent to which content is preserved during the process.
\\[0.2cm]
\textbf{Transfer Strength Accuracy. }\textit{Transfer Strength} is a metric that assesses the degree with which  the style is transferred. As proposed by Fu et al. \cite{fu2018style}, this metric is implemented using a classifier $C$. We consider that \textit{behavior expressivity style} is defined as follows:
\begin{equation}
\text{\textit{Behavior expressivity style}} = \left\{ 
\begin{array}{rcr}
\text{\textit{Source (positive) output}} \leq 0.5 \\
\text{\textit{Target (negative) output}} > 0.5 
\end{array}
\right.
\end{equation}
\textit{Transfer Strength Accuracy} is defined as follows: 
\begin{equation} \label{TSAccuracy}
	\text{\textit{Transfer Strength Accuracy}} = {\frac{N_{right}}{N_{total}} \times 100}
\end{equation}
where $N_{right}$ is the number of correct cases which are transferred from target to source style, and $N_{total}$ is the number of test set data. 
We developed $C$ as a neural network consisting of three LSTM layers and a dense output layer, with the complete architecture shown in the appendix. The network's hyperparameters were chosen empirically and are also listed in the appendix. To train $C$, we used the train sets of the speakers included in the train sets of both the $Seen$ and $Unseen$ conditions, as defined in the \textit{PATS Corpus}. Specifically, we trained $C$ using a batch size of $256$ and $Adam$ optimizer, and a \textit{Binary Cross Entropy} loss over $15,000$ epochs. After training, $C$ achieved an accuracy of $96\%$.
\\[0.2cm]
\textbf{Content Preservation. }\textit{Content Preservation} is a metric that reflects the preservation of source content, that is, in this work, the meaning conveyed by the nonverbal communicative behaviors, in predictions. It is defined as the cosine distance between predictions $\widehat{{\textit{Z}}}_{{\textit{gestures}}}$ and initial \textit{source} gestures ($X_{source}$), as follows:
\begin{equation} \label{CosineDistance}
\text{\textit{Cosine Distance}} =  1 - \frac{X_{source}^\intercal  \text { } \cdot \widehat{\textit{Z}}_{{\textit{gestures}}}}{\lVert{X_{source}\lVert}   \lVert{\widehat{Z}_{{\textit{gestures}}}}\lVert}
\end{equation}
\textbf{Minkowski distance.} 
We also measure the \textit{Minkowski distance} between the upper-body gestures and facial expressions produced by our model, and the ones of the \textit{source} and \textit{target} speakers. 
We additionally experimented with alternative distance metrics, including \textit{cityblock}, \textit{Chi2 distance}, \textit{Euclidean distance}, and \textit{cosine distance}. However, we found that all these metrics yielded identical results, leading us to retain only the Minkowski distance. More specifically, for both conditions, \textit{Seen} and \textit{Unseen}, we define two sets of distances: (1) \textbf{\emph{Dist.(}}\emph{\textit{TranSTYler}}, \emph{Source}\textbf{)} which is the average distance between \textit{TranSTYler}'s predictions and the source's 2D facial landmarks and body joint; and (2) \textbf{\emph{Dist.(}}\emph{\textit{TranSTYler}}, \emph{Target}\textbf{)} which is the average distance between \textit{TranSTYler}'s predictions and the target data.
\\[0.2cm]
\textbf{Velocity, Acceleration, Jerk. }
We evaluate the \textit{velocity}, \textit{acceleration}, and \textit{jerk} of the source, target, and prediction to quantify and compare their distinct dynamic movement patterns. This analysis enables us to determine whether the prediction aligns more closely with the behavior expressivity style of the source or of the target. Velocity provides insights into the overall speed and rhythm of the movement, while acceleration measures the rate of change in motion velocity. Jerk indicates the smoothness of motion transitions over time. By examining these metrics, we can gain valuable information about the characteristics of the movement dynamics and utilize it to assess the degree to which the predicted animation captures the behavior expressivity style of the source or target.
\subsection{Human Perceptual Studies}
\textcolor{black}{Following previous research \cite{Ahuja_CVPR_lowRes, ahuja2020style}, we contribute to the definition of a comprehensive methodology to assess behavior expressivity style transfer.  We focus on differentiating between behaviors associated with the linguistic content of speech (i.e communicative gestures), and the unique style exhibited by a speaker. The desired outcome is to preserve the gestures form associated with the source content while adjusting their expressivity to match the target style. The proposed methodology is defined as follows: 
\begin{itemize}
    \item To assess \textit{behavioral expressivity style transfer}, we evaluate the resemblance of our model's predictions to the target style.
    \item To assess \textit{content preservation}, we study the coherence of gestures by assessing their coordination with speech content and the synchronization with speech rhythm.
\end{itemize}
We conduct three studies and compare the perception of stimuli generated by our model and by the two baselines Mix-Stage and DiffGAN.}

\begin{figure}[]
\includegraphics[width=0.9\linewidth, height=6.5cm]{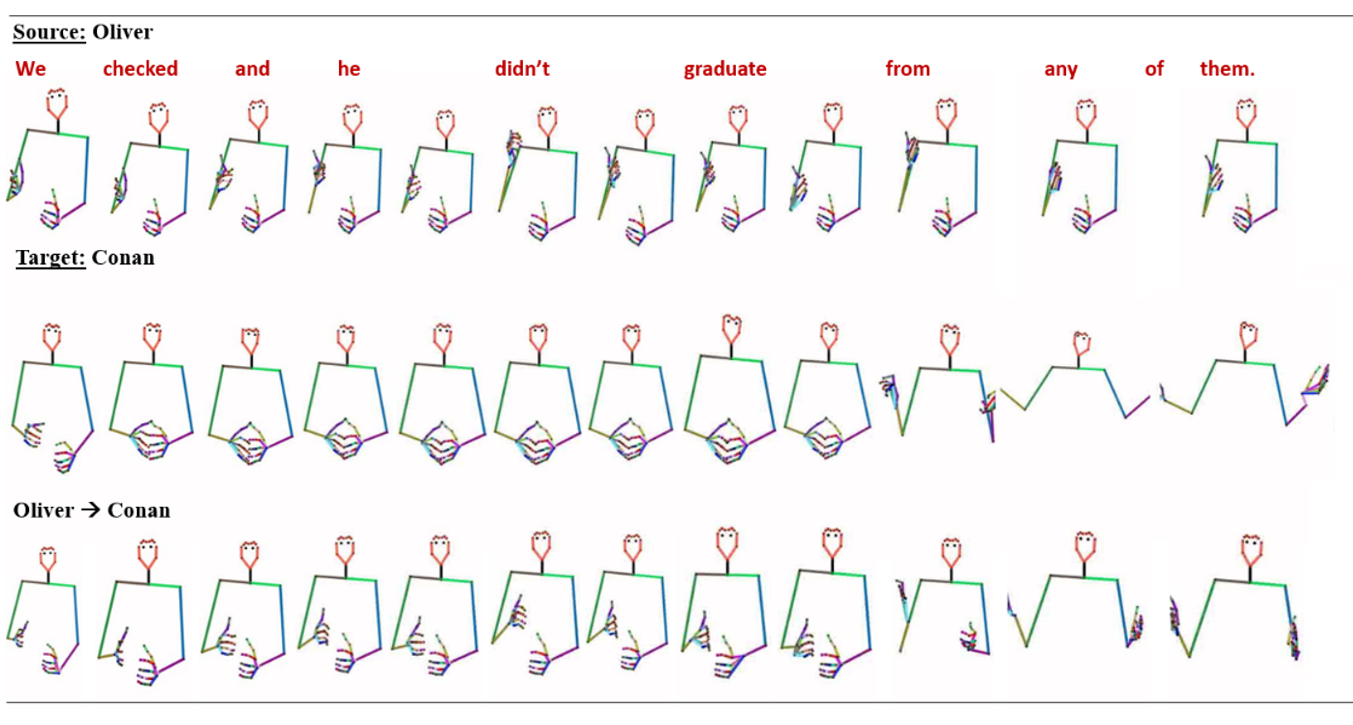}
\caption{Behavioral expressivity style transfer from target speaker Oliver to source speaker Conan. Fingers are not generated by our model but extracted from OpenPose. They are displayed for sake of visualisation.}
\label{oliver_conan}
\end{figure}

\textbf{Study 1. }The first study aims to assess the behavior expressivity style produced by our model w.r.t the behavior expressivity style of the \textit{seen} or \textit{unseen} target speakers.  We additionally compare our model to \textit{Mix-Stage}\cite{ahuja2020style}, which we consider our first baseline. We present 75 stimuli of 2D stick animation (like the 2D sticks in Fig.\ref{oliver_conan}) to evaluate our model's predictions with \textit{seen }target styles (\textit{condition 1}, 30 stimuli), our model's predictions with \textit{unseen} target styles (\textit{condition 2}, 30 stimuli), and the baseline \textit{Mix-StAGE} (\textit{condition 3}, 15 stimuli). Each stimulus consists of a triplet of 2D animations composed of: (1) a 2D animation of the source speaker, (2) a 2D animation of the target speaker, and (3) a 2D animation of \textit{TranSTYler}’s prediction after performing the behavioral style transfer. 
Participants rate on a 5-point Likert scale the \textit{overall resemblance}, \textit{resemblance of the left and right arms gesturing}, \textit{body orientation}, \textit{head orientation}, \textit{gesture amplitude}, \textit{gesture frequency}, and \textit{gesture velocity} of the target style animation with respect to the source style animation and our predictions' animation. The rating scale ranges from 1 (reference is very similar to A) to 5 (reference is very similar to B).
\\[0.2cm]
\textbf{Study 2. }We conduct a second study to investigate the coherence of the generated facial and body gestures. Previous research has focused on evaluating the appropriateness of generated gestures \cite{kucherenko2023evaluating}. 
In this work, we place greater emphasis on evaluating the coherence of gestures by assessing their coordination with speech content and synchronization with speech rhythm. By doing so, we aim to subjectively evaluate the extent to which content is preserved after performing behavioral style transfer. We evaluate the coherence of facial and body gestures in relation to speech content and rhythm. We present 90 stimuli of 2D stick animations, comprising 30 stimuli of \textit{TranSTYler}'s stylized predictions (\textit{condition 1}), 30 stimuli of \textit{TranSTYler}'s predictions where we change the original audio with the audio from other speakers (\textit{condition 2}), and 30 stimuli of the source style ground truth (\textit{condition 3}). \textit{Condition 2} is included as an error and control condition. On a 5-point Likert scale, participants rate the \textit{synchronization} of gestures with speech rhythm, the \textit{overall coherence} of behavior, the \textit{coordination} of the agent's gestures with speech content, and the \textit{human-likeness} of the animations.
\\[0.2cm]
\textbf{Study 3. }A third study is conducted to compare the similarity of our model's predictions to the target style, as well as to those generated by our second baseline, \textit{DiffGAN} \cite{Ahuja_CVPR_lowRes}. We present 15 stimuli, each comprising a triplet of 2D animations corresponding to the same source-target behavioral style transfer. The first animation is generated by \textit{TranSTYler} (\textit{condition 1}). The second animation represents the reference, and it is \textit{target speakers' ground truth}. The third animation is generated by \textit{DiffGAN} (\textit{condition 2}). For each stimulus, we ask participants to identify which video between \textit{condition 1} and \textit{condition 2} has the same behavior expressivity style as the reference video based on the arm gesture's extent, frequency, timing, and position of the body in relation to speech.
We recruited 150 participants through the online crowd-sourcing website Prolific for our evaluation studies. Participants were selected based on their fluency in English. Attention checks were included at the beginning and middle of each study to filter out inattentive participants. 
Prior to each study, participants received training to introduce them to the 2D facial landmarks and upper-body skeleton and to familiarize them with 2D stick animations. 
\section{Results and Discussion}
\begin{figure*}[]
\includegraphics[width=1\linewidth, height=7cm]{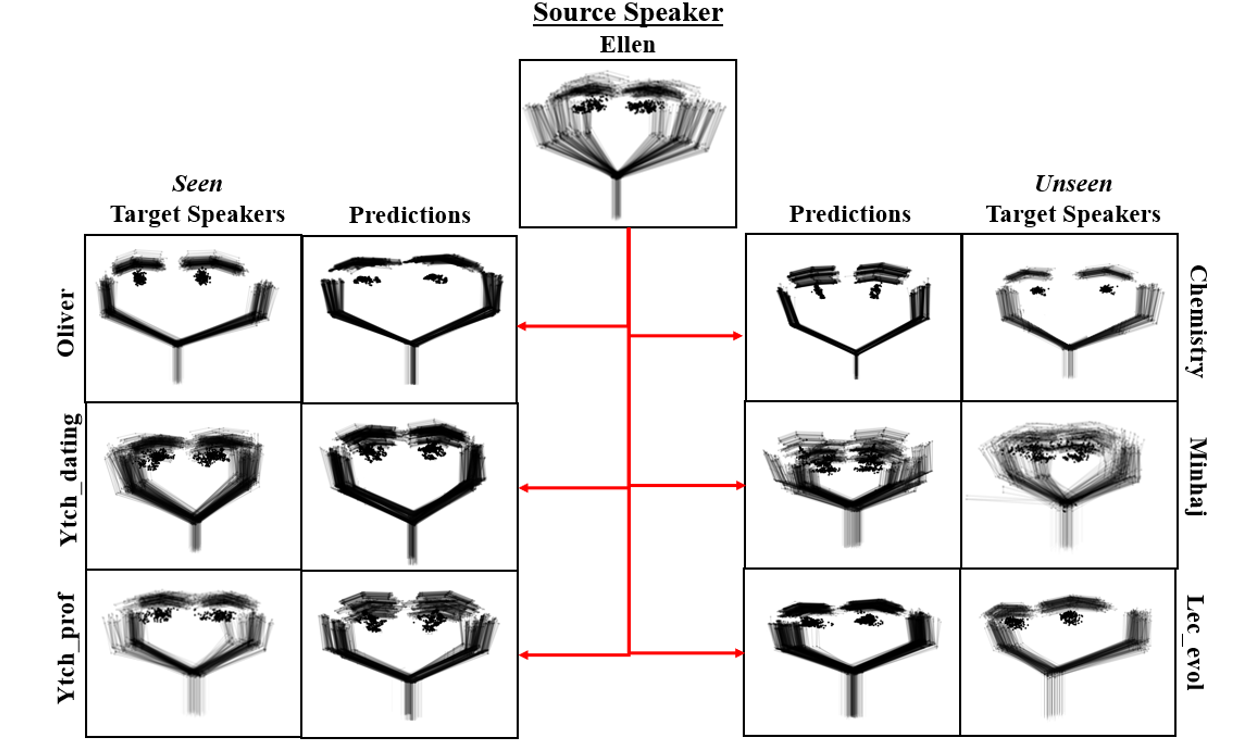}
\caption{Visualization of facial motion of source, \textit{\textit{seen}} or \textit{unseen} target and \textit{TranSTYLer}'s predictions.}
\label{face_heat}
\end{figure*}
\begin{figure*}[]
\includegraphics[width=1\linewidth, height=7cm]{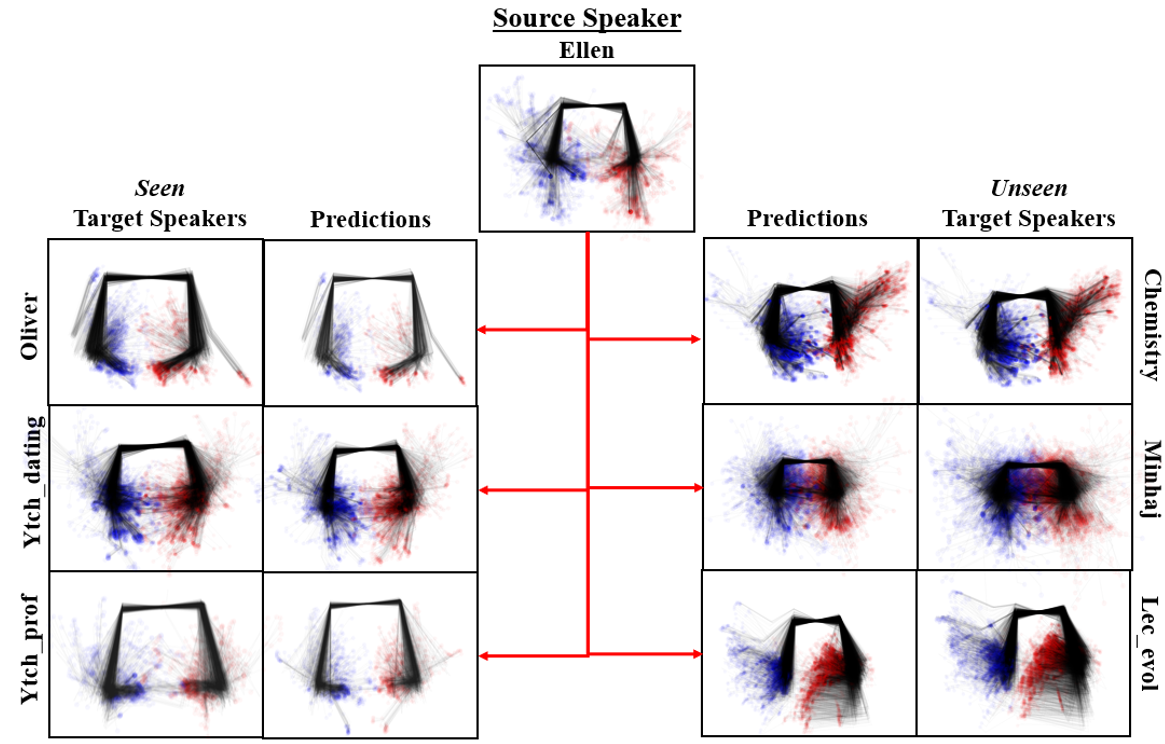}
\caption{Visualization of upper-body motion of source, \textit{\textit{seen}} or \textit{unseen} target and \textit{TranSTYLer}'s predictions.}
\label{body_heat}
\end{figure*}
\begin{figure}[]
\includegraphics[width=1\columnwidth]{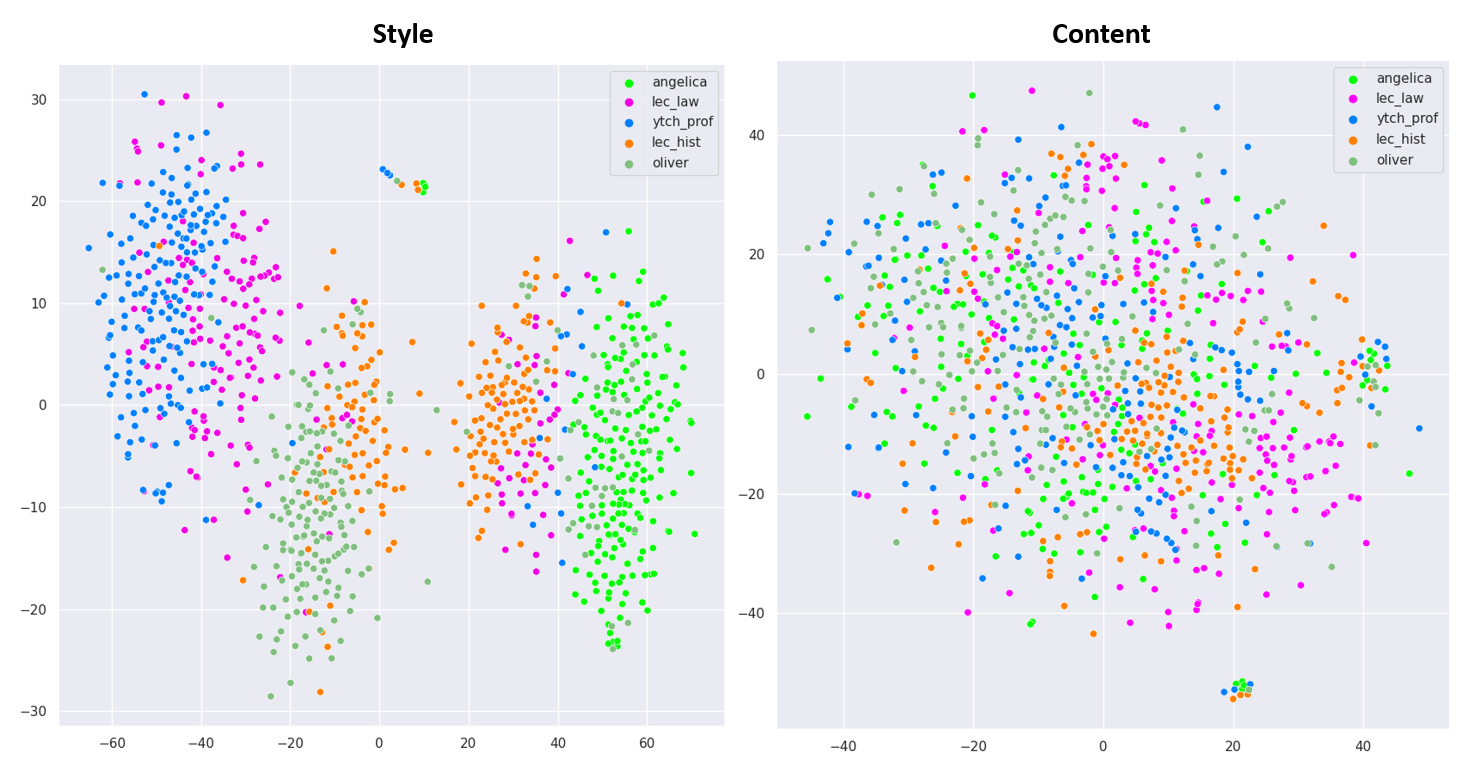}
\caption{\textbf{Left}: T-SNE visualization of the style embeddings on the test sets of 5 speakers. \textbf{Right: }T-SNE visualization of the content embeddings on the same test sets. }
\label{tsne}
\end{figure}    
\subsection{Objective Evaluation}
\begin{table}[]
\centering
\tiny
\resizebox{\columnwidth}{!}{%
\begin{tabular}{lll|ll}
\hline
\multicolumn{1}{c}{\textbf{Condition}} &
  \multicolumn{1}{c}{\textbf{\begin{tabular}[c]{@{}c@{}}Transfer Strength\\ Accuracy (\%)\end{tabular}}} &
  \multicolumn{1}{c|}{\textbf{\begin{tabular}[c]{@{}c@{}}Content \\ Preservation (\%)\end{tabular}}} &
  \multicolumn{1}{c}{\textbf{\begin{tabular}[c]{@{}c@{}}Dist.\\ w.r.t. Source\end{tabular}}} &
  \multicolumn{1}{c}{\textbf{\begin{tabular}[c]{@{}c@{}}Dist. \\ w.r.t. Target\end{tabular}}} \\ \hline \hline
\textit{Seen} &
  93.282 &
  95.842 &
  83.189 &
  75.882 \\ \hline
\textit{Unseen} &
  85.195 &
  90.723 &
  80.284 &
  73.934 \\ \hline
\end{tabular}%
}
\caption{Objective evaluation results: transfer strength accuracy, content preservation, and minkowski distances for Seen and Unseen conditions.}
\label{ObjEval}
\end{table}
\begin{figure*}[!ht]
\includegraphics[width=1.1\textwidth, height=5cm]{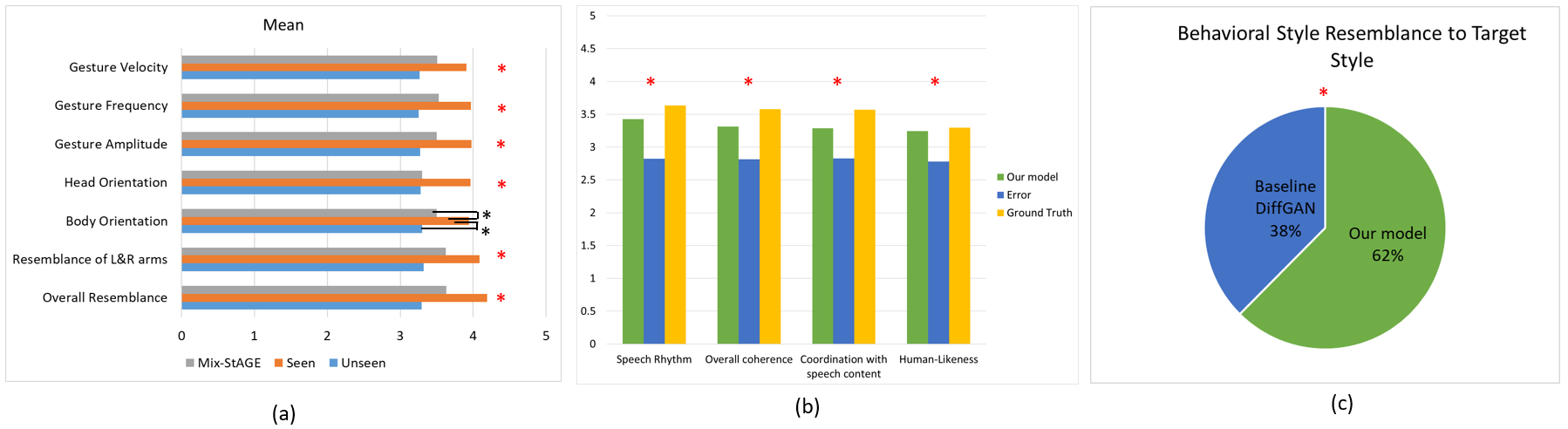}
\caption{Results of perceptual human study 1 (a), study 2 (b), and study 3 (c). Significant differences between pairs of all conditions for the same factor are marked with (\textcolor{red}{$\ast$}). If there are significant results between only two pairs of conditions for the same factor, (\textcolor{black}{$\ast$}) is used.}
\label{eval_subj}
\end{figure*}
Objective evaluation results are presented in Table \ref{ObjEval} for both \textit{Seen} and \textit{Unseen} conditions. In the \textit{Seen} condition, \textbf{\textit{TranSTYler}} achieves a style transfer strength accuracy of $93.282 \%$, indicating a high level of accuracy in transferring the style from the target speakers to the source speakers. For the \textit{Unseen} condition, the accuracy is still high at $85.195 \%$, although slightly lower than the accuracy for the \textit{Seen} condition. This was expected since \textbf{\textit{TranSTYler}} had not seen the target speakers during training. Nonetheless, the model demonstrated the ability to generalize the style to new, unseen speakers. In both the \textit{Seen} and \textit{Unseen} conditions, our model is able to preserve a high percentage of the source speakers' content, with $95.842 \%$  and $90.723 \%$ content preservation, respectively. The distance between our model's predictions and the source speakers' gestures - \textbf{\textit{Dist.(\emph{\textit{TranSTYler}}, \emph{Source})}} - is higher than the one between our model's predictions and the target speakers' gestures - \textbf{\textit{Dist.(\emph{\textit{TranSTYler}}, \emph{Target})}}. These results confirm that the \textit{behavior expressivity style} is successfully transferred from \textit{target} to \textit{source} speakers for both conditions.  These results are further illustrated in Figures \ref{face_heat} and \ref{body_heat} that illustrate the facial and body motion of a same source speaker (Ellen) as well as different target speakers that were either seen or unseen during training, alongside \textit{TranSTYler}'s predictions after performing style transfer. Figure \ref{oliver_conan} illustrates the motion of a source speaker Oliver, a target (seen) speaker Conan, and that of \textit{TranSTYler}'s predictions at the frame level. The source speaker Oliver gestures mainly with his right hand while Conan makes ample arm movements as shown toward the end of his sentence. The predicted animation displays the communicative gestures of Oliver (similar vertical movement of the arm) and the amplitude extent of Conan toward the end of the sentence. 
\textcolor{black}{To explore the relationships and patterns between the content and style vectors generated by our model, a 2D T-SNE analysis was conducted. This analysis projects the vectors onto a two-dimensional space, where their proximity indicated similarity. The TSNE plots in Figure \ref{tsne} showcase the content embeddings ($h_{content}$) and style embeddings ($h_{style}$) after disentangling the style-content information. By examining the distribution of features in the content and style space, it was observed that content and style were effectively separated. The style space exhibited clustering of features belonging to the same speaker, suggesting discernible patterns. However, in the content space, features from all speakers were mixed together without clear patterns or clusters. These results demonstrate the success of our disentangling approach in effectively separating style-content information.} We additionally computed the velocity, acceleration, and jerk of \textit{TranSTYler}'s predictions, source style, and target style for source-target style transfers where the target is either \textit{Seen} or \textit{Unseen}. The results for style transfers performed on four source-target pairs, with two \textit{Seen} targets (Angelica and Lec\_hist) and two \textit{Unseen} targets (Almaram and Minhaj) indicate that, for both \textit{Seen} and \textit{Unseen} targets, \textit{TranSTYler}'s velocity is closer to the target than to the source. 
Regarding the acceleration metric, we observed similar results for all style transfers except for the transfer from source \textbf{\textit{Bee}} to the unseen target \textbf{\textit{Minhaj}}, where the predictions' acceleration is closer to the source style. 
However, for the same \textbf{\textit{Bee - Minhaj}} style transfer, our predictions' jerkness is closer to the target than the source style. 
For the style transfer \textbf{\textit{Lec\_law - angelica}}, \textbf{\textit{TranSTYler}} produces a velocity that is close to the target style and far from the source style, an acceleration that is in between the source and target style, and a jerkness closer to the source style. Overall,  these findings show that \textit{TranSTYler} effectively transfers the behavior expressivity style from the target to the source speakers, as evidenced by the high style transfer accuracy, content preservation, and the observed patterns in velocity, acceleration, and jerk metrics.
\subsection{Subjective Evaluation}
Figure \ref{eval_subj} (a) shows the mean scores obtained for all factors for all conditions (\textit{Mix-StAGE}, \textit{Seen} and \textit{Unseen}); the higher the mean score, the closer the condition is w.r.t the target style. For all factors, our model obtained the mean scores higher than those of the baseline. For all factors, \textit{Mix-StAGE} received lower scores than the \textit{Seen} condition and higher scores than the \textit{Unseen} condition. This may be due to the fact that speakers are visible during training in the \textit{Mix-StAGE} condition, whereas \textit{TranSTYler} is unseen in the \textit{Unseen} condition. 
We conducted post-hoc paired t-tests for each factor between the three conditions and found significant differences (p < 0.007) between \textit{Mix-StAGE} and \textit{Seen}, and \textit{Unseen} and \textit{Seen} for all factors. We found significant results for \textit{Mix-StAGE} and \textit{Unseen} for all factors except 'body orientation'. Our prediction model in both \textit{Seen} and \textit{Unseen} conditions outperforms the baseline for all factors. The \textit{Seen} condition also surpasses the \textit{Unseen} one. \textcolor{black}{Our prediction model in both \textit{Seen} and \textit{Unseen} conditions outperforms the baseline for all factors. The \textit{Seen} condition also surpasses the \textit{Unseen} one.} 
The goal of Study 2 is to assess the preservation of the speech content during the style transfer. It evaluates the coherence of gestures by examining their coordination with speech content and synchronization with speech rhythm.
Results of \textbf{Study 2 } are presented in Figure \ref{eval_subj}(b). We conducted paired t-tests for each factor between the following conditions: \textit{TranSTYler} and \textit{Error}, \textit{TranSTYler} and \textit{Ground Truth}, and \textit{Ground Truth} and \textit{Error}. The results showed significant differences ($p<0.001$) between each pair of conditions for all factors. \textit{TranSTYler} achieved scores that are significantly ($p<0.001$) very similar to the ground truth scores. In contrast, the control condition, \textit{error}, obtained a significantly ($p<0.001$) lower mean score than the scores obtained by our model. Thus the gestures computed by our model maintained adequacy with the speech content as predicted gestures are highly similar to those in the original video. However, we are aware that further study ought to be conducted on measuring more precisely the semantic and pragmatic information conveyed by the predicted behaviors. 
The third study aimed to compare our model, \textit{TranSTYler}, with a second baseline, \textit{DiffGAN}\cite{Ahuja_CVPR_lowRes}, and results are shown in Figure \ref{eval_subj} (c). Participants were asked to identify which animation, between condition 1 (\textit{TranSTYler}) and condition 2 (\textit{DiffGAN}), had the most similar behavior expressivity style as the reference video (\textit{target style}) based on the arm gesture's extent, frequency, timing, and position of the body in relation to speech. 
A post-hoc binomial test was also conducted, and significant results were found for both conditions ($p<0.001$). Thus, overall, our model generates animations that significantly capture better the behavior expressivity style of the target speaker than does \textit{DiffGAN}.
\section{Conclusion}
We present \textit{TranSTYLer} for synthesizing body and facial gestures of source speakers in the style of target speakers, without additional training. We propose a novel methodology for evaluating behavior expressivity style transfer, measuring content preservation and style transfer while identifying potential leakages between style and content information. Furthermore, we expand the PATS corpus by including 2D Facial Landmarks and Dialog Tags.

\bibliographystyle{ACM-Reference-Format}
\bibliography{references} 

\end{document}